\documentclass[twocolumn,preprintnumbers,amsmath,amssymb,showpacs]{revtex4}
\usepackage{epsfig}
\begin{document}
\setlength{\voffset}{1.0cm}
\title{From non-degenerate conducting polymers to dense matter \\ in the massive Gross-Neveu model}
\preprint{FAU-TP3-05/4}
\author{Michael Thies}
\author{Konrad Urlichs\footnote{Electronic addresses:
thies@theorie3.physik.uni-erlangen.de,  konrad@theorie3.physik.uni-erlangen.de}}
\affiliation{Institut f\"ur Theoretische Physik III,
Universit\"at Erlangen-N\"urnberg, D-91058 Erlangen, Germany}
\date{\today}
\begin{abstract}
Using results from the theory of non-degenerate conducting polymers like {\em cis}-polyacetylene,
we generalize our previous work on dense baryonic matter and the soliton crystal in the massless
Gross-Neveu model to finite bare fermion mass. In the large $N$ limit, the exact crystal ground state 
can be constructed analytically, in close analogy to
the bipolaron lattice in polymers. These findings are contrasted to the standard scenario with 
homogeneous phases only and a first order phase transition at a critical chemical potential.  
\end{abstract}
\maketitle
\section{Introduction}
In this paper we continue our investigation of the Gross-Neveu (GN) model \cite{1}, a 1+1 dimensional, asymptotically free 
relativistic quantum field theory with Lagrangian
\begin{equation}
{\cal L} = \bar{\psi}^{(i)} ({\rm i}\gamma^{\mu}\partial_{\mu} - m_0)\psi^{(i)} + \frac{1}{2} g^2 (\bar{\psi}^{(i)}\psi^{(i)})^2.
\label{A1}
\end{equation}
The index $i$ runs over $N$ fermion species, and we are only concerned with the 't Hooft limit $N\to \infty$, $Ng^2=$ const.
If the bare fermion mass $m_0$ vanishes, this model has a discrete chiral symmetry $\psi\to \gamma^5 \psi$.
During the last few years, it has become clear that the phase diagram at finite temperature and chemical 
potential of the massless GN  model is richer than previously thought \cite{2}, the most striking novel 
feature being the appearance of a solitonic crystal phase.
This phase diagram can be computed exactly, with many results in closed analytical form \cite{3,4}. The
corresponding analysis for the massive case ($m_0 \neq 0$) has yet
to be done. As a first step in this direction, we have recently determined the properties of single baryons in the
massive GN model \cite{5}. Here we address the problem of dense baryonic matter at zero temperature
in the massive GN model.

Interestingly, the GN model has lived a kind of double life in particle physics and condensed matter physics
over many years, with only sporadic cross-references (see e.g.~\cite{6,7,8}). On the one hand, it is a soluble model
for strong interaction physics, exhibiting phenomena like asymptotic freedom, chiral symmetry breaking, 
dynamical mass generation,
meson and baryon bound states \cite{9}. On the other hand,
it can model quasi-one dimensional systems in the vicinity of a half-filled band, in particular conducting polymers \cite{9a}. 
A prominent example is {\em trans}-polyacetylene (PA). This polymer (CH)$_x$ possesses two dimerized, degenerate
ground states (Fig.~1) and, owing to a number of simplifying assumptions, leads to a continuum description mathematically
equivalent to the symmetric ($m_0=0$) GN model.
\begin{figure}
\epsfig{file=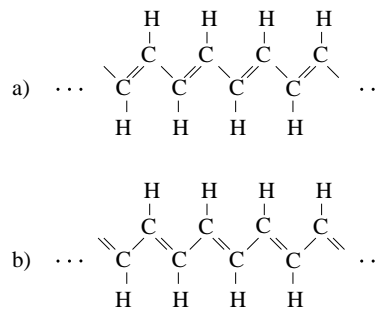,width=5cm}
\caption{Two degenerate, dimerized ground states of {\em trans}-PA, described theoretically in terms
of the massless GN model.}
\end{figure}
Dimerization, i.e., the fact that short and long bonds alternate, plays the role of (discrete) chiral symmetry breaking \cite{9b}.
Solitons and polarons are important for the physics of doping and electrical conductivity. They have originally been found
by analogy with kink and kink-antikink baryons already known on the field theory side \cite{10}. By and large, the
analysis of the phase
diagram has evolved independently in the two fields and only recently merged into a consistent picture. 

If one starts to think about the massive GN model ($m_0\neq 0$), it is not too hard to identify its 
condensed matter analogue: Conducting polymers with non-degenerate ground states
like {\em cis}-PA (Fig.~2).
\begin{figure}
\epsfig{file=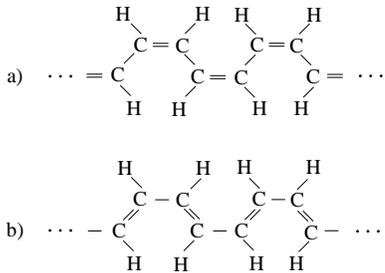,width=5cm}
\caption{Two inequivalent configurations of {\em cis}-PA, leading to a description in terms of 
the massive GN model. Configuration a) has lower energy than b).}
\end{figure}
Their theoretical description was initiated by Brazovskii and Kirova \cite{11} with the proposal
that the gap parameter has two contributions, a constant, ``external"  one arising from the basic structure of the 
polymer and an $x$-dependent, ``internal" one due to electron-phonon coupling,
\begin{equation}
\Delta (x) = \Delta_e + \Delta_i(x)  .
\label{A2}
\end{equation} 
If we identify $\Delta(x)$ with the scalar mean field $S(x)$ and $\Delta_e$ with the bare mass $m_0$, we can immediately 
relate two problems from two different branches of physics. Baryons in the massive GN model correspond to
polarons and bipolarons in the polymer case \cite{5}. In the 80's and 90's, a lot of work was devoted to
conducting polymers \cite{12} with the result that exact bipolaron lattice solutions were found by several authors.
This strongly suggests that the massive GN model should exhibit a soliton crystal at finite baryon density, much
like the massless model. 
One has to be careful though, since in detail the use of the GN model is different in the two fields. 
In the relativistic quantum field case, one has to send the UV cutoff to infinity and the bare parameters 
$Ng^2$ and $m_0$ to 0 in a specific way, dictated by the renormalizability of the model. In polymer physics,
the role of the UV cutoff is taken over by a band width $W$, a physical parameter (incidentally of the order of 10 eV).
Likewise, the electron-phonon coupling $\lambda$  and the external gap parameter $\Delta_e$ are physical observables
to be taken from experiment. Nevertheless, in practice
these differences do not seem to matter for many questions. We have noticed that condensed matter calculations
are often done using approximations tantamount to the standard renormalization procedure in field theory.
A similar remark applies to the role of $N$, the number of fermion species: Although $N$ is as small as 2  
in polymer physics (the electron spin components), the adiabatic
calculations done are in most cases indistinguishable from the large $N$ treatment on the field theory side. 

In the present work, we explore the possibility that the ground state of the massive GN model at finite baryon
density has a crystal structure akin to the bipolaron crystal in polymers. 
We proceed as follows: We take advantage of the fact that the functional form of the periodic
self-consistent scalar potential for non-degenerate polymers is known. It has been derived by various methods
such as inverse spectral theory \cite{13}, Poisson summation of periodic sums of single polarons \cite{14,15,16} or 
relation to the Toda lattice \cite{7}. (Ref.~\cite{17} contains a particularly thorough discussion and corrects
misprints in some of the original papers.) We shall use the two-parameter functional form determined in these works as
ansatz in a relativistic Hartree-Fock (HF) calculation, see Eqs.~(\ref{B5},\ref{B7}) below, but otherwise perform an
independent calculation. 

Our main motivation is to generalize our recent calculation of the phase diagram of the massless GN model
to $m_0 \neq 0$. The most detailed investigation of the phase diagram of the massive GN model 
to date is the work of Barducci et al.  \cite{18}.
These authors assume that the chiral condensate is spatially constant and derive the phase
diagram for finite temperature and chemical potential. At $T=0$ in particular, they observe a 1st order phase 
transition at a critical chemical potential depending on the bare fermion mass. Judging from what is known about the
 massless limit,
these findings can probably not be trusted, since the existence of the baryon lattice has not yet been taken into account. 

Equipped with the structure of the single baryon in the massive GN model, the soliton crystal in   
the massless limit and a candidate for the self-consistent potential in the massive case borrowed from polymer physics,
we should have all the ingredients necessary to clarify this problem. In the present work, we
concentrate on finite baryon density at $T=0$ only.
In Sec.~II, we present the calculation of the ground state energy, assuming a periodic scalar potential
of a definite functional form. Sec.~III is devoted to the proof that the
ansatz is indeed general enough to yield self-consistency, together with the determination of its free parameters.
In Secs.~II and III, we rely on some results from Ref.~\cite{3} where the same problem 
has been solved in the chiral limit, but nevertheless try to keep the presentation reasonably self-contained.  
Since the formulae are fairly complicated and their derivation is involved, we shall spend some time checking
our results in various limits: The chiral limit $m_0 \to 0$ in Sec.~IV, the low density limit in Sec.~V and the high-density limit
in Sec.~VI. Once the formulae have passed all of these tests, we turn to exhibit their physics content and present some 
illustrative numerical results in Sec.~VII. We summarize what we have learned in the concluding section, Sec.~VIII.  Two
appendices
are devoted to the definitions of elliptic integrals and functions (Appendix A) and to the translationally invariant solution
of the massive GN model, needed here primarily to assess the stability of the soliton crystal (Appendix B).

\section{Ground state energy of baryonic matter in the Hartree-Fock approach}

Before describing in detail our calculations we give a short overview of what will be done next.
In the present section, we shall compute the ground state energy density of baryonic matter in the relativistic
HF approach. We assume right away a periodic ansatz for the scalar potential $S(x)$.
This ansatz is borrowed from condensed matter physics and contains two free parameters, see Eqs.~(\ref{B5},\ref{B7}).
It generalizes the scalar potential given in Eq.~(6) of Ref.~\cite{3} which was successful in the chiral limit
of the GN model. The important feature of the ansatz which will enable us to carry out the calculation analytically is
the fact that it yields the simplest, single gap Lam\'e potential in the equivalent
Schr\"odinger-type equation, just like in the chiral limit. Therefore many technicalities of the present calculation
are unchanged as compared to this limit and a number of results from Ref.~\cite{3} can be taken over directly. The
main difference 
between the two calculations arises from the modified relation between the Dirac-HF single
particle energies and the spectrum of the Lam\'e equation. This renders 
summing up single particle energies of occupied orbits somewhat tedious in the massive model. 
Although the final result of this section can still be given in closed analytical form, it requires incomplete elliptic
integrals of all three kinds (see Appendix A) as opposed to complete elliptic integrals in the chiral limit.
This result will be presented in the form $E_{\rm g.s.}=E_1+E_2$ in Eqs.~(\ref{B19},\ref{B24}),
where $E_1$ is the sum over single particle energies of occupied states and $E_2$ the double counting correction
to the interaction energy characteristic for the HF approach. It generalizes 
Eq.~(24) in Ref.~\cite{3} to which it reduces in the chiral limit. 
The final expression for $E_{\rm g.s.}$ of the present section still contains
two undetermined parameters introduced through the ansatz for the scalar potential. The crucial question is then
whether the ansatz is general enough to encompass the self-consistent potential. This will be answered
positively in Sec.~III.

The first steps in the computation of the ground state energy density are the same as in the chiral limit. We set up the 
Dirac-HF equation which for the GN model (\ref{A1}) reads
\begin{equation}
\left( \gamma^5 \frac{1}{\rm i} \frac{\partial}{\partial x} + \gamma^0 S(x) \right)\psi_{\alpha}(x)=\omega_{\alpha} \psi_{\alpha}(x).
\label{B1}
\end{equation}
The scalar potential obeys the self-consistency equation which follows from minimizing the 
ground state energy, $\delta E_{\rm g.s.}/\delta S(x)=0$, 
\begin{equation}
- \frac{1}{Ng^2}(S(x)-m_0)  = \sum_{\alpha}^{\rm occ.}\bar{\psi}_{\alpha}(x) \psi_{\alpha}(x).
\label{B1a}
\end{equation}
The sum runs over all occupied single particle states up to the Fermi energy, including
the filled negative energy states of the Dirac sea. 
In the present section, we concentrate
on evaluating the ground state energy for our ansatz for the scalar potential. The proof of the
self-consistency condition (\ref{B1a}) which is technically the most difficult part of the whole
calculation is deferred to Sec.~III. 

From previous work \cite{18} it is known that the HF problem posed in Eqs.~(\ref{B1},\ref{B1a}) has a solution with
a uniform scalar potential $S(x)=M$ (a physical fermion mass). For the convenience of the reader we have
sketched this ``standard" solution in Appendix B. There we also cover the vacuum case and show how
to replace the dimensionless coupling $g$ and the bare mass $m_0$  by physical parameters, thereby defining
our renormalization procedure. The relation between bare and physical parameters [see Eqs.~(\ref{B21},\ref{B22})] will be 
crucial to cancel divergences from the Dirac sea, for instance on the right hand side of Eq.~(\ref{B1a}). 
Apart from these well understood issues, the translationally invariant approach involves only straightforward
calculations, since all we have to do is to solve a free, massive Dirac equation. 
By contrast, a periodic ansatz for the scalar potential $S(x)$ makes the HF problem much more
challenging. Fortunately, we will see that owing to the specific form of $S(x)$ the problem 
remains tractable.

Returning to Eq.~(\ref{B1}), we use the following representation of the $\gamma$-matrices,
\begin{equation}
\gamma^0 = - \sigma_1 , \quad \gamma^1={\rm i}\sigma_3, \quad \gamma^5=\gamma^0\gamma^1 = - \sigma_2.
\label{B2}
\end{equation}
In terms of upper and lower spinor components $\phi_{\pm}$ (suppressing the orbit label $\alpha$ from hereon), we
obtain the pair of coupled equations
\begin{equation}
\pm \left( \frac{\partial}{\partial x} \mp S \right) \phi_{\mp} = \omega \phi_{\pm} 
\label{B3}
\end{equation}
which can be decoupled by squaring
\begin{equation}
\left( - \frac{\partial^2}{\partial x^2} \mp \frac{\partial S}{\partial x} + S^2\right) \phi_{\pm} = \omega^2 \phi_{\pm}.
\label{B4}
\end{equation}
We now use the ansatz
\begin{equation}
S(x)=A\tilde{S}(\xi) \qquad  (\xi=Ax)
\label{B5}
\end{equation}
where $\tilde{S}$ is of the general functional form taken from work on bipolaron crystals in non-degenerate
polymers \cite{13,14,15,16,17}
\begin{equation}
\tilde{S} = \frac{{\rm sn}\,\xi_+ {\rm cn}\, \xi_+ {\rm dn}\, \xi_+ + {\rm sn}\,\xi_- {\rm cn}\,\xi_- {\rm dn}\,\xi_-}
{{\rm sn}^2\xi_+-{\rm sn}^2\xi_-}.
\label{B7}
\end{equation}
Our conventions and shorthand notation for elliptic integrals and Jacobian elliptic functions are summarized
in Appendix A, and 
\begin{equation}
\xi_{\pm} = \xi \pm \delta.
\label{B8}
\end{equation} 
As is well known from quasi-one-dimensional condensed matter systems, the physics behind the appearance
of periodic structures is the Peierls instability --- the system lowers its energy by generating dynamically 
an energy gap at the Fermi surface \cite{18a}. This was already observed in the chiral limit of the GN model
where the state of lowest energy has a completely filled (for matter) or empty (for antimatter) valence band.
Assuming that the gap is located at the Fermi surface, the scale factor $A$ takes on the same value as in the chiral limit
[$\bf{K}$ is the complete elliptic integral of the first kind, Eq.~(\ref{J4})], 
\begin{equation}
A= \frac{2 p_f {\bf K}}{\pi},
\label{B6}
\end{equation}
with the Fermi momentum $p_f$ related to the spatially averaged baryon density via $\rho=p_f/\pi$. 
The two real parameters which determine $S(x)$ are the elliptic modulus $\kappa$, suppressed in the notation of
Eqs.~(\ref{B7},\ref{B6}), and the shift $\delta$. 
With the ansatz (\ref{B7}), Eq.~(\ref{B4}) yields the simplest form of the Lam\'e equation
\begin{equation}
\left(- \frac{\partial^2}{\partial \xi^2}+ 2 \kappa^2 {\rm sn}^2 \xi_{\mp} \right) \phi_{\pm} = {\cal E} \phi_{\pm},
\label{B8a}
\end{equation}
laying the ground for an analytical solution. The relation between Dirac-  and Lam\'e eigenvalues
is 
\begin{equation}
{\cal E} = \left( \frac{\omega^2}{A^2}-\eta \right), \quad \omega=\pm A\sqrt{{\cal E}+\eta},
\label{B9}
\end{equation}
where $\eta$ is found to be 
\begin{equation}
\eta  = \frac{1}{{\rm sn}^2 2\delta}  -1 - \kappa^2.
\label{B10}
\end{equation}
The spectrum and eigenfunctions of the Lam\'e equation (\ref{B8a}) are well-known \cite{18b}. The energy
bands $\kappa^2 \leq {\cal E} \leq 1$ and $1+\kappa^2 \leq {\cal E}$ are separated by a single gap, whereas 
the density of states inside the allowed bands is given by \cite{18c}
\begin{equation}
\left| \frac{{\rm d}k}{{\rm d}{\cal E}} \right| = \frac{\left|{\bf E}/{\bf K} + \kappa^2-{\cal E}\right|}{2 \sqrt{(1-{\cal E})
({\cal E}- \kappa^2)(1+ \kappa^2 -  {\cal E})}}.
\label{B13}
\end{equation}
Note that Eq.~(\ref{B8a}) is identical to what we had at $m_0=0$; at this point differences arise from
the relation between Dirac single particle energies $\omega$ and Lam\'e eigenvalues ${\cal E}$, Eqs.~(\ref{B9},\ref{B10}),
and from reconstructing solutions of the original first order equations (\ref{B3}) where the shape of the potential $S(x)$ enters.
At a later stage, $m_0$ also appears in the self-consistency equation, see Eq.~(\ref{B1a}) and Sec.~III.

In the present section, we focus on the ground state energy density
\begin{equation}
E_{\rm g.s.} = E_1 + E_2,
\label{B11}
\end{equation}
where $E_1$ is the sum (or rather integral) over single particle energies of all filled 
orbits (including the negative energy states in the Dirac sea) and $E_2$ the HF double-counting correction.
Proceeding like at $m_0=0$, we first sum over single particle energies,
\begin{equation}
E_1 = -2 A^2 \int_{{\cal E}_<}^{{\cal E}_>} \frac{{\rm d}{\cal E}}{2\pi} \left| \frac{{\rm d}k}{{\rm d}{\cal E}}\right|
\sqrt{{\cal E}+\eta},
\label{B12}
\end{equation}
with the integration limits  ($\Lambda/2$ is the UV cutoff) 
\begin{eqnarray}
{\cal E}_< & = & 1 + \kappa^2,
\nonumber \\
{\cal E}_> & = & \left( \frac{\Lambda}{2A}\right)^2  + 2 \left( 1-{\bf E}/{\bf K} \right)
\label{B13a}
\end{eqnarray}
[{\bf E} is the complete elliptic integral of the second kind, Eq.~(\ref{J5})]. 
We consider antimatter here, filling only the negative energy states below the gap. 
This yields the same ground state energy as for matter, albeit in a simpler way.
The integration in Eq.~(\ref{B12}) using (\ref{B13}) is significantly more involved that in the massless case, but can be done
along similar lines as in Ref.~\cite{14}.
The result is best expressed in terms of incomplete elliptic integrals $F,E,\Pi$ of the first, second and third kind
(see Appendix A) as follows,
\begin{eqnarray}
\frac{2\pi}{A^2} E_1 & = & -  \sqrt{\frac{({\cal E}_>-\kappa^2)({\cal E}_>-1-\kappa^2)({\cal E}_>+\eta)}
{({\cal E}_>-1)}}
\nonumber \\
& & +   \chi \left( 2 \frac{\bf E}{\bf K} - 2 + \kappa^2 \right) F(p,q) +  \chi E(p,q) 
\nonumber \\
& & +   \frac{\kappa^2}{\chi} \left( 2 \frac{\bf E}{\bf K} - 2 - \eta\right)
\Pi(p,n,q),
\label{B14}
\end{eqnarray}
where we have introduced
\begin{equation}
\chi=\sqrt{1+\eta} = \frac{{\rm dn}\, 2\delta}{{\rm sn}\, 2 \delta}
\label{B14a}
\end{equation}
for a recurring expression to ease the notation. The arguments of the elliptic
integrals are defined as follows,
\begin{eqnarray}
n& =& \frac{1+ \kappa^2+\eta}{1+\eta} \ = \ \frac{1}{{\rm dn}^2 2\delta}
\nonumber \\
p &=& \sqrt{\frac{{\cal E}_> - 1 - \kappa^2}{n ({\cal E}_>-1)}}
\nonumber \\
q & = & \sqrt{n(1-\kappa^2)} \ = \ \frac{\kappa'}{{\rm dn}\, 2\delta}
\label{B15}
\end{eqnarray}
($\kappa'=\sqrt{1-\kappa^2}$ is the complementary elliptic modulus). Unlike in condensed matter physics, we
are only interested in the limit $\Lambda\to \infty$, so that we only need the asymptotic behavior
of $E_1$ for large ${\cal E}_>$.
The first term in Eq.~(\ref{B14}) yields an irrelevant quadratic divergence $-\Lambda^2/(8\pi)$ which can be dropped,
as well as a finite term. In the 2nd and 3rd terms, we can replace $p$ by $\tilde{p}$ in $E(p,q)$ and $F(p,q)$, where
\begin{equation}
\tilde{p} = \frac{1}{\sqrt{n}}= {\rm dn}\, 2\delta ,
\label{B16}
\end{equation} 
all corrections being $1/{\cal E}_>$ suppressed. 
The last term in Eq.~(\ref{B14}) is more delicate, since $\Pi(p,n,q)$ has a logarithmic singularity at $p=\tilde{p}$. 
Expanding
\begin{equation}
p \approx \frac{1}{\sqrt{n}}- \frac{\kappa^2}{2 \sqrt{n} {\cal E}_>} = \tilde{p} - \epsilon
\label{B17}
\end{equation}
and using a standard identity \cite{19} in the form
\begin{eqnarray}
\Pi(\tilde{p}-\epsilon,n,q) &=&   \frac{1}{2} \sqrt{\frac{n}{(n-1)(n-q^2)}}\ln \frac{2(n-q^2)(n-1)}{\sqrt{n}(n^2-q^2)\epsilon}
\nonumber \\
&+& F(\tilde{p},q)    - \Pi\left(\tilde{p},\frac{q^2}{n},q\right)  + {\rm O}(\epsilon),
\label{B18} 
\end{eqnarray}
we finally arrive at  
\begin{eqnarray}
\frac{2\pi}{A^2} E_1 & = &  \left( 2 \frac{\bf E}{\bf K} - 2 + \kappa^2- \frac{1}{2} \eta\right)
\nonumber \\
& + &  \left( 2 \frac{\bf E}{\bf K} - 2 -\eta\right) \ln \frac{\Lambda}{A\sqrt{2+\eta}} +    \chi E(\tilde{p},q) 
\nonumber \\
& + &  \frac{\kappa^2}{\chi}\left(2\frac{\bf E}{\bf K} -2 - \eta \right) \left[F(\tilde{p},q)
-\Pi\left(\tilde{p},(\kappa')^2,q\right)\right]
\nonumber \\
& + &  \chi \left( 2 \frac{\bf E}{\bf K} - 2 + \kappa^2\right) F(\tilde{p},q).
\label{B19}
\end{eqnarray}
The logarithmic divergence of the sum over single particle energies (the $\ln \Lambda$ term in the 2nd line) 
will be cured once we add the double counting correction $E_2$, to which we now turn.
For finite bare quark mass $m_0$ it is given by ($\ell =2 {\bf K}$ is the spatial period in $\xi$),
\begin{equation}
E_2 = \frac{1}{2 Ng^2} \frac{1}{\ell} \int_0^{\ell} {\rm d}\xi (A \tilde{S}(\xi)-m_0)^2 .
\label{B20}
\end{equation}
We invoke the vacuum gap equation [see Appendix B, Eq.~(\ref{I2})] to eliminate the bare coupling constant, 
\begin{equation}
\frac{1}{Ng^2} \approx \frac{1}{\pi} (1+ m_0)\ln \Lambda  .
\label{B21}
\end{equation}
We use natural units in which the dynamical fermion mass in the vacuum is $m=1$.
In the massive GN model, there are two physical, renormalization group invariant parameters,
the fermion mass $m$ and 
\begin{equation}
\gamma = m_0 \ln \Lambda.
\label{B22}
\end{equation} 
[Notice that our definition of $\gamma$ (called ``confinement parameter" in the
condensed matter literature) differs by a factor of $\pi$ from the definition in Refs.~\cite{5,20}. It turns out
that the present definition simplifies the notation below.] We then find
\begin{equation}
E_2 = \frac{A^2}{2\pi\ell}\left( \gamma +  \ln \Lambda \right) \int_0^{\ell} {\rm d}\xi \tilde{S}^2(\xi)
-  \frac{A\gamma}{\pi \ell} \int_0^{\ell} {\rm d}\xi \tilde{S}(\xi). 
\label{B23}
\end{equation}
Performing the integrations analytically, the final result for the double-counting correction becomes 
\begin{eqnarray}
E_2 & = &- \frac{A^2}{2\pi} \left(\gamma + \ln \Lambda\right )\left( 2 \frac{\bf E}{\bf K} -2 - \eta \right),
\nonumber \\
 &  & +   \frac{A\gamma}{\pi}  \left( f_1 - f_2   \frac{{\bf \Pi}(\kappa^2 {\rm sn}^2 \delta, \kappa)}{\bf K} \right).
\label{B24} 
\end{eqnarray}
with 
\begin{eqnarray}
f_1 & = & \frac{{\rm cn}^2 \delta + {\rm dn}^2 \delta + {\rm cn}^2 \delta {\rm dn}^2 \delta}
{2\, {\rm sn}\,\delta\, {\rm cn}\,\delta\, {\rm dn}\,\delta},
\nonumber \\
f_2 & = & \frac{2 \,{\rm cn}\,\delta \,{\rm dn}\,\delta}{{\rm sn}\,\delta}
\label{B25}
\end{eqnarray}
[${\bf \Pi}$ is the complete elliptic integral of the 3rd kind, Eq.~(\ref{J6})]. 
 
Eqs.~(\ref{B19}) and (\ref{B24}) are the main result of this section. Upon adding up $E_1$ and $E_2$ to
get the ground state energy density $E_{\rm g.s.}$, the logarithmically divergent terms are cancelled and a finite
result involving only physical parameters is obtained. It depends on the 4 parameters 
$\kappa, \delta, p_f, \gamma$. Out of these 4 parameters, $p_f$ and $\gamma$ are determined 
by the baryon density and the bare fermion mass, respectively, whereas $\kappa$ and $\delta$
are so far unspecified parameters of the trial potential. They will be determined in the
following section by demanding self-consistency.

\section{Self-consistency condition}
In the present section we check the self-consistency of the ansatz scalar potential Eqs.~(\ref{B5},\ref{B7}). 
This ansatz contains two free parameters, $\kappa$ and $\delta$. We proceed to show 
that the self-consistency condition (\ref{B1a}) is satisfied provided that $\kappa$ and $\delta$ take on definite values
depending on the density and the parameter $\gamma$.
The main result of this section will be a pair of transcendental equations for $\kappa$ and $\delta$, Eqs.~(\ref{C19}),
which are equivalent to the self-consistency equation. 

The self-consistency condition for the scalar potential, Eq.~(\ref{B1a}), can be rewritten as
\begin{equation}
 - \frac{S}{\pi}(\gamma + \ln \Lambda) + \frac{\gamma}{\pi}
 =   \sum_{\alpha}^{\rm occ.}\bar{\psi}_{\alpha}(x) \psi_{\alpha}(x),
\label{C3}
\end{equation}
We have used Eqs.~(\ref{B21}) and (\ref{B22}) to trade the bare parameters $Ng^2$ and $m_0$ against $\gamma$
and the cutoff $\Lambda$.
Our next task is therefore to evaluate the scalar density $\bar{\psi}\psi$ for an arbitrary single particle solution
of the Dirac-HF equation (\ref{B1}) and perform the sum over all occupied states.
Since the first step involves only properties of the Lam\'e eigenfunctions and is practically identical to
the corresponding calculation in the appendix of Ref.~\cite{3}, we skip some straightforward 
intermediate steps and just state the result,
\begin{equation}
\bar{\psi}\psi = \frac{A \kappa^2}{2\omega \left( {\rm dn}^2\alpha - {\bf E}/{\bf K}\right)}
 \left(\partial_{\xi}+ 2 \tilde{S}\right) \left( {\rm cn}^2 \alpha  - 
{\rm cn}^2 \xi \right).
\label{C6}
\end{equation}
The parameter $\alpha$ parametrizes the spectrum of the Lam\'e Hamiltonian, see Eq.~(\ref{C10}) below.
For the present purpose, it is actually more convenient to shift the variable $\xi$
by $\delta$ and decompose the potential $\tilde{S}$, Eq.~(\ref{B7}), into 
two terms,
\begin{equation}
\tilde{S} = \tilde{S}_1+\tilde{S}_2
\label{C7}
\end{equation}
with
\begin{eqnarray}
\tilde{S}_1 & = & \frac{{\rm sn}\,\xi_{++}{\rm cn}\,\xi_{++}{\rm dn}\,\xi_{++}}{{\rm sn}^2\xi_{++}-{\rm sn}^2\xi},
\nonumber \\
\tilde{S}_2 & = & \frac{{\rm sn}\,\xi\,{\rm cn}\,\xi\, {\rm dn}\,\xi}{{\rm sn}^2\xi_{++} - {\rm sn}^2\xi}.
\label{C8}
\end{eqnarray}
We have introduced the doubly shifted argument
\begin{equation}
\xi_{++} = \xi+2\delta.
\label{C2}
\end{equation}
The contribution to the condensate from a single orbit is then given by
\begin{equation}
\bar{\psi}\psi = \frac{A\kappa^2}{\omega \left({\rm dn}^2\alpha - {\bf E}/{\bf K}\right)}
\left( - \tilde{S} {\rm sn}^2 \alpha + \tilde{S}_1 {\rm sn}^2\xi + \tilde{S}_2 {\rm sn}^2\xi_{++} \right).
\label{C9}
\end{equation}
Using 
\begin{eqnarray}
{\rm dn}^2 \alpha & = & {\cal E}- \kappa^2 \nonumber \\
{\rm sn}^2 \alpha & = & \frac{1}{\kappa^2} \left(1-{\cal E}+ \kappa^2 \right)
\label{C10}
\end{eqnarray}
and summing over occupied states (as in Sec.~II we consider antimatter) then yields
\begin{eqnarray}
\sum_{\alpha}^{\rm occ.} \bar{\psi}_{\alpha} \psi_{\alpha}&=& 2 A \int_{{\cal E}_<}^{{\cal E}_>}
\frac{{\rm d}{\cal E}}{2\pi} \left| \frac{{\rm d}k}{{\rm d}{\cal E}} \right| \bar{\psi}\psi
\label{C11} \\
& = & - \frac{A}{2\pi} \left[ I_2 \tilde{S}+ \kappa^2 I_1 ( \tilde{S}_1 {\rm sn}^2\xi  + \tilde{S}_2 {\rm sn}^2\xi_{++})\right]
\nonumber 
\end{eqnarray}
with the integrals  
\begin{eqnarray}
I_1 & = & \frac{2}{\chi} F(p,q),
\nonumber \\
I_2 & = &   \frac{2\kappa^2}{\chi} \left[ \Pi(p,n, q) -F(p,q) \right].
\label{C12}
\end{eqnarray}
The arguments of the elliptic integrals are the same as above, see Eqs.~(\ref{B15}). Using once again
the identity (\ref{B18})
to isolate the logarithmic singularity of $\Pi(p,n,q)$ and taking the asymptotics with respect to $\Lambda$
allows us to replace
\begin{eqnarray}
F(p,q) & \to & F(\tilde{p},q) ,
\nonumber \\
\Pi(p,n,q) & \to &  F(\tilde{p},q)  - \Pi\left(\tilde{p},(\kappa')^2,q\right)
\nonumber \\
& & + \frac{\chi}{\kappa^2} \ln \frac{\Lambda}{A\sqrt{2+\eta}}.
\label{C13}
\end{eqnarray}
Upon inserting the results (\ref{C11}--\ref{C13}) into the self-consistency condition (\ref{C3}), 
the $\ln \Lambda$ term drops out once again and we arrive at the finite equation
\begin{eqnarray}
\frac{\chi \gamma}{A\kappa^2}(S-1)  & = & 
 F(\tilde{p},q )\left( \tilde{S}_1 {\rm sn}^2\xi + \tilde{S}_2 {\rm sn}^2\xi_{++} \right) 
\label{C14} \\
&- &  \left[ \Pi\left(\tilde{p},(\kappa')^2,q\right) + \frac{\chi}{\kappa^2} \ln (A\sqrt{2+\eta})\right]\tilde{S}.
\nonumber
\end{eqnarray}
To clarify the content of this condition, we re-arrange it slightly and introduce coefficients ${\cal C}_{1,2}$
as follows,
\begin{eqnarray}
 \gamma & = & {\cal C}_1 \left[ \tilde{S} + \frac{{\cal C}_2}{{\cal C}_1}  
\left( \tilde{S}_1{\rm sn}^2\xi + \tilde{S}_2 {\rm sn}^2\xi_{++} \right) \right],
\nonumber \\
{\cal C}_1 & = & A \ln (A \sqrt{2+\eta}) + A \gamma  + \frac{A \kappa^2}{\chi}
 \Pi\left(\tilde{p},(\kappa')^2,q\right),
\nonumber \\
{\cal C}_2 & = & - \frac{A \kappa^2}{\chi} F(\tilde{p},q).
\label{C15}
\end{eqnarray}
At first glance, it seems unlikely that Eq.~(\ref{C15}) can be solved, since a constant term and 
two different functions of $x$ appear.
However, these functions are not linearly independent. 
Using addition theorems for cnoidal functions, one can verify the following identity,
\begin{equation}
\tilde{S}- \kappa^2 {\rm sn}^2 2\delta \left( \tilde{S}_1 {\rm sn}^2\xi  + \tilde{S}_2 {\rm sn}^2\xi_{++} \right)
= \frac{{\rm cn}\,2\delta\, {\rm dn}\,2\delta}{{\rm sn}\,2\delta}.
\label{C16}
\end{equation}
Comparing Eqs.~(\ref{C15}) and (\ref{C16}),  
the self-consistency condition may be turned into the following two $x$-independent conditions,
\begin{eqnarray}
\frac{\gamma}{{\cal C}_1} & = & \frac{{\rm cn}\,2\delta \,{\rm dn}\,2\delta}{{\rm sn}\,2\delta},
\nonumber \\
\frac{{\cal C}_2}{{\cal C}_1} & = & - \kappa^2 {\rm sn}^2 2\delta,
\label{C20}
\end{eqnarray}
with ${\cal C}_{1,2}$ as defined in Eq.~(\ref{C15}).
For $\gamma \neq 0$, these equations can be cast into the somewhat simpler form
\begin{eqnarray}
0 & = & A\, {\rm cn}\,2\delta  F(\tilde{p},q) -  \gamma\,  {\rm sn}^2 2\delta,
\label{C19} \\
\gamma & = & A \, {\rm cn}\,2\delta  \left[ \kappa^2 \Pi\left(\tilde{p},(\kappa')^2,q\right) + \chi
\left( \gamma  + \ln (A \sqrt{2+\eta}) \right) \right].
\nonumber
\end{eqnarray}
They determine $\kappa$ and $\delta$ for given $p_f$ and $\gamma$.

After carefully identifying all the variables, the self-consistency equations (\ref{C19}) agree
exactly with Eqs.~(5,6) in \cite{17}, confirming the 1:1 mapping from the theory of 
non-degenerate conducting polymers to the massive GN model. 
If we can solve the pair of transcendental equations (\ref{C19}), we have found a solution
of the HF equations and hence a candidate for the ground state of baryonic matter. 
Another solution of the HF equations is the translationally invariant one
discussed in detail in Ref.~\cite{18} and summarized in Appendix B.
Which solution is favored is then simply a question of 
the energy which can be computed from Eqs.~(\ref{B19},\ref{B24}) of Sec.~II.
The numerical results will be presented in Sec.~VII. The following three sections are 
devoted to testing the general formalism in simple special cases where we know the 
answer from other sources.

\section{Chiral limit}

As a first and most trivial test, let us check the above formalism in the chiral limit ($\gamma=0$) against our previous work.
Consider the self-consistency equation, Eq.~(\ref{C20}). The first equation can be solved by 
\begin{equation}
\delta = \frac{\bf K}{2} , \qquad {\rm cn}\, 2\delta = 0.
\label{D1}
\end{equation}  
Consequently,
\begin{equation}
\tilde{p}=\chi=\kappa',\quad q=1, \quad \eta=-\kappa^2,
\label{D2}
\end{equation}
and the elliptic integrals are reduced to elementary functions,
\begin{eqnarray}
F(\kappa',1) & = & {\rm artanh}\,  \kappa' ,
\label{D3} \\
\Pi(\kappa',(\kappa')^2,1) & = & \frac{1}{2\kappa^2} \left( \kappa' \ln \frac{\kappa^2}{1+(\kappa')^2}+\ln \frac{1+\kappa'}
{1- \kappa'} \right).
\nonumber 
\end{eqnarray}
The second equation of Eqs.~(\ref{C20}) then yields simply
\begin{equation}
A\kappa=1,
\label{D4}
\end{equation}
relating $\kappa$ and $p_f$. Likewise, the
ground state energy Eqs.~(\ref{B19},\ref{B24}) simplifies tremendously in the limit $\gamma \to 0$,
\begin{equation}
E = \frac{A^2}{4\pi} \left( 4 \frac{\bf E}{\bf K}-2 + \kappa^2 \right)
- \frac{A^2}{2\pi} \left( 2 \frac{\bf E}{\bf K} -2 + \kappa^2 \right) \ln A\kappa.
\label{D5}
\end{equation}
Finally, starting from Eqs.~(\ref{C7},\ref{C8})  and using standard identities for
cnoidal functions of the shifted argument $\xi+{\bf K}$, the self-consistent potential
can be manipulated into the form
\begin{equation}
\tilde{S} = \kappa^2 \frac{{\rm sn}\, \xi\,  {\rm cn}\, \xi}{{\rm dn}\, \xi}.
\label{D6}
\end{equation}
All of these formulae agree with the known results for $\gamma=0$ \cite{3}. 
A comparison of the full equations of Secs.~II and III with the reduced ones in the present section 
shows a significant (but apparently unavoidable) increase in complexity due to the finite bare fermion mass.

\section{Low density limit and single baryon}
A second obvious way of testing the formalism is the low density limit, where we expect to
recover the properties of single baryons in the massive GN model. We have to expand
$\kappa=\sqrt{1-(\kappa')^2}$ around $\kappa'=0$. To leading order,
we may neglect all power corrections in $\kappa'$ but must keep the logarithmic singularity
$\sim \ln \kappa'$ whenever it is present. In this way, we find 
\begin{equation}
\tilde{p}  \approx {\rm sech} \, 2 \delta,  \qquad q \approx  0,
\label{E1}
\end{equation}
and hence all incomplete elliptic integrals reduce to the same elementary function,
\begin{equation}
\left\{ F(\tilde{p},q),E(\tilde{p},q),\Pi(\tilde{p},(\kappa')^2,q)\right\}\to \arcsin \tilde{p}.
\label{E2}
\end{equation}
The self-consistency conditions (\ref{C19}) become 
\begin{eqnarray}
0 &=& 2 p_f  \tilde{p}\arcsin \tilde{p} \ln \frac{4}{\kappa'}    - \pi \gamma \tanh^2 2\delta
\label{E3} \\
\pi \gamma  & = & 2 p_f \tilde{p} \ln  \frac{4}{\kappa'} \left[ \arcsin \tilde{p} +\frac{1}{\sinh 2\delta}
\left(\gamma + \Gamma\right) \right] 
\nonumber
\end{eqnarray}
with the definition
\begin{equation}
\Gamma = \ln \left( \frac{2p_f \ln(4/\kappa')}{\pi \tanh 2\delta} \right).
\label{E4}
\end{equation}
These equations can be solved parametrically for $\delta$ and $\kappa'$ as follows,
\begin{eqnarray}
\delta  &=& \frac{1}{2}{\rm artanh}\, y ,
\nonumber \\
\kappa' &=& 4 \exp \left\{ - \frac{\pi y}{2 p_f}\right\},
\label{E5}
\end{eqnarray}
where
\begin{eqnarray}
y & = & \sin \theta,
\nonumber \\
0 & = & \frac{\pi}{2}- \theta - \gamma \tan \theta,
\label{E6}
\end{eqnarray}
in perfect agreement with the corresponding equations for the single baryon 
(with completely filled valence level) in the massive GN model \cite{5}.

Using our full expression for the energies $E_1$ and $E_2$, Eqs.~(\ref{B19},\ref{B24}), the limit $\kappa' \to 0$ yields
the finite parts
\begin{eqnarray}
E_1 & = & - \frac{1}{4\pi} + \frac{p_f}{\pi} \left( \frac{2y\gamma}{\pi} + \frac{2y}{\pi}\right),
\nonumber \\
E_2 & = & -\frac{\gamma}{2\pi} + \frac{p_f}{\pi} \left( \frac{2y}{\pi} + \frac{2\gamma}{\pi} {\rm artanh}\,y\right).
\label{E8}
\end{eqnarray}
The logarithmically divergent pieces have been omitted, since they anyway cancel in the sum. 
The total energy density can then be represented as
\begin{equation}
E_{\rm g.s.} = E_1+E_2 = E_{\rm vac} + \rho M_B
\label{E9}
\end{equation}
where the vacuum energy density agrees with Eq.~(\ref{I4}) in the appendix, $\rho=p_f/\pi$ is the mean baryon
density and $M_B$ is the known baryon mass \cite{5}. This is indeed the
expected low density behavior for widely spaced baryons and an important additional test
of the formalism. 

Finally, we check the shape of the scalar potential in the low density limit. Taking $\kappa \to 1$ in Eq.~(\ref{B7}), we find
\begin{equation}
\lim_{\kappa \to 1} \tilde{S}(\xi) = \coth 2 \delta - \frac{\sinh 2 \delta}{\cosh^2 \xi + \cosh^2 \delta -1}. 
\label{E10}
\end{equation}
Let us compare this expression to the profile of the single baryon at $m_0 \neq 0$,
\begin{eqnarray}
S_B(x) &=& 1 + y \left[ \tanh(yx-c_0)-\tanh(yx+c_0)\right],
\nonumber \\
c_0 &=& \frac{1}{2} {\rm artanh}\, y .
\label{E11}
\end{eqnarray}
If we identify
\begin{equation}
A=y, \quad \xi=yx  ,\quad \delta = c_0 ,
\label{E12}
\end{equation}
this can be replaced by 
\begin{equation}
\tilde{S}_B(\xi) = \coth 2 \delta + \tanh (\xi-\delta) - \tanh(\xi + \delta)
\label{E13}
\end{equation}
in agreement with Eq.~(\ref{E10}).
One can easily convince oneself that the relations (\ref{E11},\ref{E12}) are equivalent to the 
self-consistency equations (\ref{E5}).

\section{High density limit and perturbation theory}

In the high density limit $p_f \to \infty$, one would expect that all interaction effects can be treated perturbatively. As
is well known, band formation in a periodic potential requires almost degenerate perturbation theory (ADPT). Here we 
closely follow a similar calculation carried out by us in the context of the massless GN model \cite{2}.
There are two differences: In the double-counting correction, we have to
take into account the bare quark mass, and we allow for $S_0 \neq 0$ in addition to $S_{\pm 1}\neq 0$
[$S_{\ell}$ are the Fourier components of the periodic potential $S(x)$].
Thus our present ansatz for $S(x)$ is 
\begin{equation}
S(x)=S_0 + S_1{\rm e}^{{\rm i} 2p_f x} + S_{-1} {\rm e}^{-{\rm i} 2p_f x}
\label{F1}
\end{equation}
The potential has a spatial period $a$ equal to the inter-baryon distance, i.e., the inverse density
\begin{equation} 
a=\frac{\pi}{p_f}.
\label{F1a}
\end{equation}
This is the  reason why the lowest non-vanishing momentum which appears in the
Fourier expansion (\ref{F1}) is $2\pi/a=2 p_f$.
Without loss of generality, we may assume that $S_0$ and $S_1=S_{-1}$ are real (a phase in $S_1$ corresponds to
a translation).
The sum over single particle energies (for the antimatter case) in 2nd order perturbation theory (PT) is given by
\begin{eqnarray}
E_1 &=& 2 \int _{p_f}^{\Lambda/2} {\rm d}k \left\{ -k - \frac{S_0^2}{2k} - \frac{S_1^2}{2(k+p_f)}\right.
\nonumber  \\
& & \left. - \sqrt{(k-p_f)^2+S_1^2} + (k-p_f)\right\}
\nonumber \\
& = & - \frac{\Lambda^2}{8\pi}  + \frac{p_f^2}{2\pi} - \frac{S_1^2}{4\pi} 
- \frac{1}{2\pi} \left(S_0^2+2 S_1^2\right)\ln \frac{\Lambda}{2}
\nonumber \\
& &  + \frac{1}{2\pi}(S_0^2+S_1^2) \ln p_f + \frac{S_1^2}{2\pi}\ln S_1.
\label{F2}
\end{eqnarray} 
We had to invoke  ADPT only for the term which would blow up in naive 2nd order PT,
\begin{equation}
-\frac{S_1^2}{2(k-p_f)} \to - \sqrt{(k-p_f)^2+S_1^2} + (k-p_f)
\label{F3}
\end{equation}
This ``recipe" has been derived in Eq.~(2.11) of Ref.~\cite{2} by simply comparing 2nd order degenerate
and non-degenerate perturbation theory. In Ref.~\cite{2}, the states which are almost degenerate
in a periodic potential are explained in more detail, following the standard weak binding approximation
from solid state physics (see e.g.~\cite{20a}). 
The double-counting correction for the potential (\ref{F1}) becomes
\begin{eqnarray}
E_2 & = & \frac{1}{2Ng^2} \frac{1}{L} \int_0^L {\rm d}x (S(x)-m_0)^2
\nonumber \\
& = & \frac{1}{2\pi}(S_0^2+2 S_1^2) (\gamma + \ln \Lambda) - \frac{\gamma}{\pi} S_0.
\label{F4}
\end{eqnarray} 
Adding Eqs.~(\ref{F2}) and (\ref{F4}) then yields the approximate energy density  
\begin{eqnarray}
E_{\rm g.s.} & = & - \frac{\Lambda^2}{8\pi} + \frac{p_f^2}{2\pi} + \frac{S_0^2}{2\pi} \ln (2p_f)
+ \frac{\gamma}{2\pi} \left( S_0^2-2 S_0 \right)
\nonumber \\
& & - \frac{S_1^2}{2\pi} + \frac{S_1^2}{2\pi} \ln (4 p_f S_1) + \frac{\gamma}{\pi} S_1^2.
\label{F5}
\end{eqnarray}
Minimizing with respect to $S_0$ and $S_1$, we find  
\begin{eqnarray}
S_0\left[\ln(2 p_f)+ \gamma\right]- \gamma = 0,
\nonumber \\
S_1\left[ 2 \gamma + \ln (4 p_f S_1)\right]=0.
\label{F6}
\end{eqnarray}
The first equation has the unique solution
\begin{equation}
S_0 = \frac{\gamma}{\gamma + \ln (2p_f)}
\label{F7}
\end{equation}
in agreement with the leading term in Eq.~(\ref{I12}) of the appendix. 
The 2nd equation has two solutions: $S_1=0$, corresponding to unbroken translational invariance
as discussed in Appendix B, and
\begin{equation}
S_1= \frac{1}{4 p_f}{\rm e}^{-2\gamma}
\label{F8}
\end{equation}
for the soliton crystal. Comparing the energy densities of these two solutions,
\begin{equation}
E_{\rm g.s.}(S_1\neq 0) - E_{\rm g.s.}(S_1=0) = - \frac{1}{64\pi p_f^2} {\rm e}^{-4 \gamma}
\label{F9}
\end{equation}
we learn that the crystal is favored, but the energy difference decreases rapidly with increasing $\gamma$.
Eqs.~(\ref{F8}) and (\ref{F9}) agree with our previous results if, in addition to the high density limit, 
we take the chiral limit $\gamma \to 0$. 

\section{Numerical results}
\begin{figure}
\epsfig{file=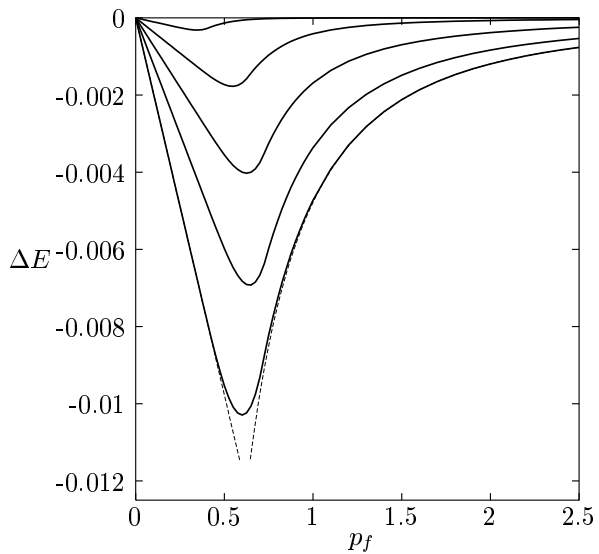}
\caption{Difference in energy density between crystal and translationally invariant Fermi gas. From top to
bottom: $\gamma=2.3,0.75,0.3,0.1,0.01$. Dashed curves: asymptotic behavior according to Eqs.~(\ref{E9}) and (\ref{F9}), for
$\gamma=0.01$.}
\end{figure}
Before computing any observable, we have to solve the self-consistency equations (\ref{C20}). 
Choosing the mean density and the bare fermion mass, or, equivalently, $p_f$ and $\gamma$, these
two transcendental equations yield the two unknown parameters $\kappa$ (elliptic modulus)
and $\delta$ (shift parameter) of the trial potential defined in Eqs.~(\ref{B5}--\ref{B6}). Eqs.~(\ref{C20})
always have two
different solutions, a translationally invariant one ($\kappa=1$) and a spatially modulated one ($0<\kappa<1$). 
In the case of the translationally invariant solution, there is an additional complication already familiar from the
massless limit: At low densities, a mixed phase appears, characteristic of a first order phase transition.
Therefore, one cannot simply take the HF energy at face value, but has to introduce a second variational
parameter describing the amount of space filled with ``droplets" of baryonic matter. 
Alternatively, one can work with a chemical potential and use the grand canonical potential at $T=0$ as
in Ref.~\cite{18}, leading to the same results. 
More details are given in Appendix B. The energy density of the crystal phase can be computed by inserting
the self-consistent values for $\kappa$ and $\delta$ into Eqs.~(\ref{B19},{\ref{B24}). We then find that
the crystal is energetically favored at all densities and all bare quark masses. In the low and
high density limits, this can be shown analytically, see Eqs.~(\ref{I23},\ref{I24}) for the limit $p_f \to 0$
and Eq.~(\ref{F9}) for $p_f \to \infty$. In between, we have to compute the energy difference
numerically (where ``numerically" in this context simply means using floating point commands in Maple, thus getting 
any desired accuracy). Some illustrative
results are shown in Fig.~3 for five different values of $\gamma$. Together with the lowest curve
(corresponding to $\gamma=0.01$),
we have also plotted two dashed curves corresponding to the (analytical) asymptotic behavior at small 
and large densities. We observe that the agreement of the full calculation with the asymptotic curves
is excellent, with a narrow crossover region where the curve changes rather abruptly
from $\sim p_f$ to $\sim p_f^{-2}$ behavior.

On the basis of such calculations, we conclude that the lattice solution is stable at all densities and quark masses. In polymer
physics, the situation is somewhat different, and stability of the bipolaron lattice has been found in certain
regions of parameter space only \cite{17}. The difference can be traced back to the fact that in the condensed
matter case, the cutoff $\Lambda$ is not sent to infinity, but is a physical parameter (the bandwidth of the
undimerized polymer). Here the GN model is used as an ``effective theory", like the Nambu-Jona-Lasinio
model in particle physics. One therefore has more parameters to vary, see e.g.~Fig.~4 of Ref.~\cite{17}. The
horizontal axis in this plot is labelled $\Delta_e/\bar{\Delta}$, corresponding to $m_0/m$ in our case. Since the bare
fermion mass goes to 0 in the limit $\Lambda \to \infty$, we can only compare our calculation with
 the point $\Delta_e/\bar{\Delta}=0$,
where there is indeed stability also in the polymer case.
The unstable region for finite values of $m_0$ is simply not accessible if one treats the GN model strictly as a
renormalizable relativistic quantum field theory.

Having established the stability of the crystal solution, let us now illustrate how a finite bare quark mass modifies
the self-consistent scalar potential. This is exhibited in Figs.~4 and 5 at low and high density, respectively.
\begin{figure}
\epsfig{file=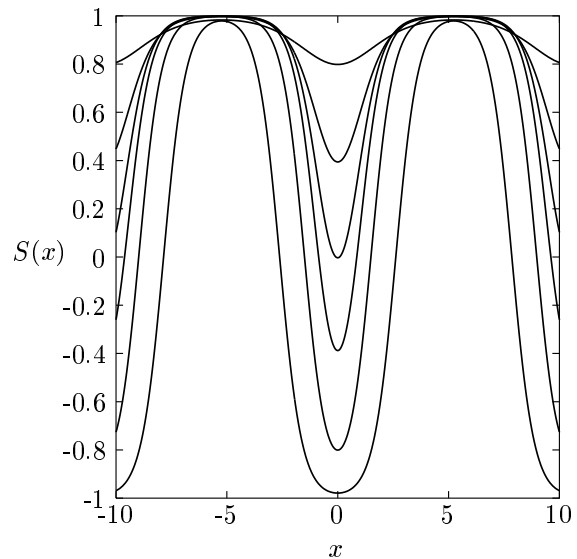}
\caption{Self-consistent scalar potential $S(x)$ versus $x$ for $p_f=0.3$. From bottom to top:
$\gamma= 0,0.01,0.1,0.3,0.75,2.3$.}
\end{figure}
\begin{figure}
\epsfig{file=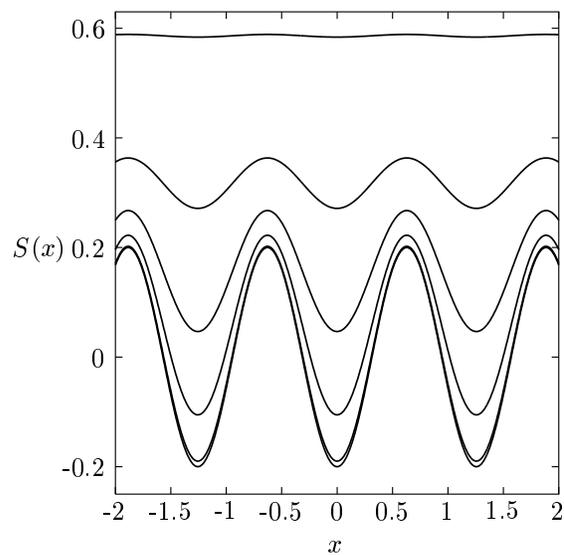}
\caption{Same as Fig.~4, but for $p_f=2.5$. Note the different scale on the $x$-axis.}
\end{figure}
The deepest curves always correspond to $\gamma=0$, where the potential oscillates symmetrically around zero.
This is actually a remnant of the original discrete chiral symmetry of the model. Translational invariance and the $\gamma^5$
transformation both break down, leaving an unbroken discrete symmetry (translation by half a period, combined with a
$\gamma^5$ transformation). Since the massive GN model does not have discrete chiral symmetry in the first place,
one would not expect the same behavior here. Indeed, the potentials now wiggle around a finite value, with a less
symmetric shape.
In the massive case, continuous translational invariance is broken down to a residual discrete translational invariance.
As we turn on the symmetry violation parameter $\gamma$, the potential oscillates with 
decreasing amplitude around a value close to the mass $M$ which the fermions would acquire 
in the translationally invariant
solution, eventually leaving only a very weak modulation of a large scalar potential in the heavy fermion limit.
It is surprising that such a variety of potential shapes in the Dirac equation can all be reduced to the 
standard single gap Lam\'e equation.

The last result which we should like to show is how the density varies with the chemical potential. The chemical 
potential at $T=0$ can be obtained by differentiating the energy density with respect to the mean fermion density,
\begin{equation}
\mu =   \frac{\partial E_{\rm g.s.}}{\partial \rho}, \qquad \rho=\frac{p_f}{\pi}.
\label{G1}
\end{equation} 
If we assume unbroken translational invariance (Fig.~6), we find discontinuities in these curves, confirming the result
of Ref.~\cite{18} about a first order phase transition. Repeating the same calculation for the crystal solution (which is the
stable one), all the curves become continuous, signalling a 2nd order phase transition (Fig.~7).   
The critical chemical potential in this latter case coincides with the baryon mass, as expected on general grounds.
By contrast, the first order transition in Fig.~6 happens at a chemical potential which has at best the meaning of an
approximate baryon mass in a kind of droplet model, cf.  Appendix B. 
\begin{figure}
\epsfig{file=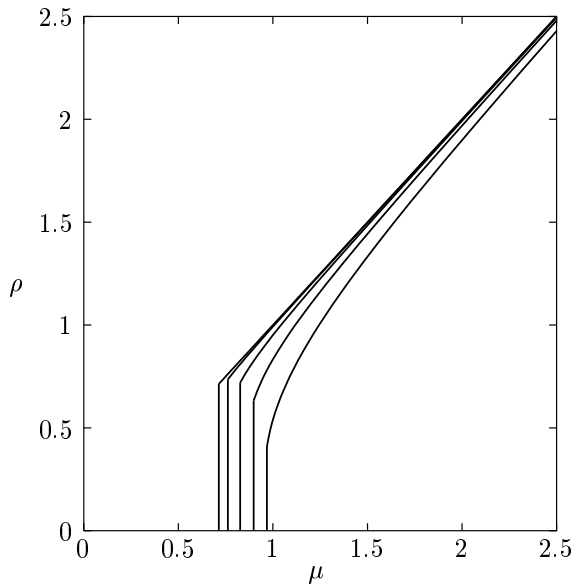}
\caption{Baryon density versus chemical potential for translationally invariant solution, showing a first order transition.
From left to right: $\gamma=0.01,0.1,0.3,0.75,2.3$.}
\end{figure}
\begin{figure}
\epsfig{file=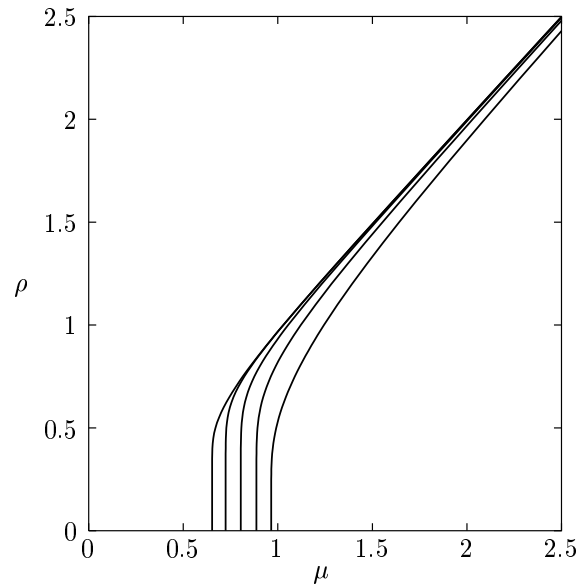}
\caption{Same as Fig.~6, but for crystal solution. Here, a continuous, second order phase transition at $\mu_c=M_B$ is seen for
all values of $\gamma$.}
\end{figure}

\section{Summary and conclusions}
In this paper, we have derived analytically the ground state of cold, dense matter in the massive GN model. This task
was greatly simplified by the lucky circumstance that the problem is mathematically closely related to another problem
which had already been solved in condensed matter physics, namely the bipolaron lattice in non-degenerate
conducting polymers. Since we could not be sure that subtle differences between the polymer problem and 
relativistic quantum field theory don't matter, we only took over the functional form of the self-consistent potential
from literature and carried out an independent HF calculation with this two-parameter ansatz. Our results 
fully confirm the 1:1 mapping between two seemingly unrelated physics problems. Whereas a similar 
relationship had been noticed before by several authors in the context of degenerate polymers and the 
massless GN model, this time we were able to take full advantage of the correspondence in a situation where
we could never have guessed the correct potential.

As compared to our previous work on the massless GN model, the calculations become significantly more complex
once one turns on the fermion mass parameter $\gamma$. In order to write down the ground state energy, one now needs 
all 6 standard complete and incomplete elliptic integrals. The results
exhibit a wider variety of scalar potentials oscillating not around 0, but around a finite value. Since the Dirac equation 
with all of these potentials can still be reduced to the single gap Lam\'e equation, the calculations remain tractable.
The crystal is energetically favored over the translationally invariant solution at all densities and quark masses
as a result of the Peierls effect with gap formation at the Fermi surface.
Like in the chiral limit, the alleged first order phase transition discussed in the literature disappears in favor
of a 2nd order phase transition. The critical chemical potential now coincides with the baryon mass, an important
theoretical constraint which had been violated by the translationally invariant solution. 
It would be very interesting to extend this work to finite temperature and determine the full phase
diagram of the massive GN model. What role does the soliton crystal play at finite temperature, and what happens
to the first order transition line ending at a critical point \cite{18}? Provided that the ansatz for
the scalar potential is flexible enough to describe the finite temperature case as well, this would be rather
straightforward, following Ref.~\cite{4} for the massless limit. As an additional bonus, such a 
calculation might even be of some interest for polymer physics.

Having seen the quantitative correspondence between polymers and the GN model, one wonders whether other
results could be traded between the Phys.  Rev.  D and Phys.  Rev.  B communities and might shed some new light
on the respective physics questions. One such issue might be asymptotic
freedom, the property to which the GN model owes its existence in the first place. Is there anything analogous
to signals of asymptotic freedom (like scaling in deep inelastic scattering) in polymer physics, perhaps in optical
properties of conducting polymers? Conversely, in the condensed matter literature, the symmetry breaking
parameter $\gamma$ is known as ``confinement parameter". Since the ground state is no longer degenerate, 
the effective kink-antikink potential rises linearly with distance. 
Could this mean that the GN model can teach us anything about confinement in quantum chromodynamics?
It will be interesting to follow further such speculations and identify questions where an intensified
dialogue between particle and condensed matter physicists might be fruitful. 
\vskip 0.5cm
We should like to thank Oliver Schnetz for mathematical support and useful conversations.
We also thank the referee of Physical Review D for helpful suggestions of how to improve the
readability of an earlier version of this paper.

\appendix
\section{Elliptic integrals and Jacobi elliptic functions}
Since different ways of writing down elliptic integrals and Jacobi elliptic functions are in use, we
briefly summarize our conventions. As a rule, we use the Legendre normal form for elliptic integrals. Our notation
of the arguments is the same as in Maple.
\begin{itemize}
\item Incomplete elliptic integral of the first kind
\begin{equation}
F(z,\kappa) = \int_0^z {\rm d}t \frac{1}{\sqrt{1-t^2}\sqrt{1-\kappa^2 t^2}}
\label{J1}
\end{equation}
\item Incomplete elliptic integral of the second kind
\begin{equation}
E(z,\kappa) = \int_0^z {\rm d} t  \frac{\sqrt{1-\kappa^2 t^2}}{\sqrt{1-t^2}}
\label{J2}
\end{equation}
\item Incomplete elliptic integral of the third kind
\begin{equation}
\Pi(z,n,\kappa) = \int_0^z {\rm d}t \frac{1}{(1-nt^2)\sqrt{1-t^2}\sqrt{1-\kappa^2 t^2}}
\label{J3}
\end{equation}
\item Complete elliptic integral of the first kind
\begin{equation}
{\bf K} = {\bf K}(\kappa) = F(1,\kappa)
\label{J4}
\end{equation}
\item Complete elliptic integral of the second kind
\begin{equation}
{\bf E} = {\bf E}(\kappa) = E(1, \kappa)
\label{J5}
\end{equation}
\item Complete elliptic integral of the third kind
\begin{equation}
{\bf \Pi}(n,\kappa) = \Pi (1,n,\kappa) 
\label{J6}
\end{equation}
\end{itemize}
In Jacobi elliptic functions, we suppress the elliptic modulus $\kappa$ throughout this paper,
\begin{equation}
{\rm sn}\,u = {\rm sn}(u, \kappa), \  \ {\rm cn}\,u = {\rm cn}(u, \kappa), \ \   
{\rm dn}\,u = {\rm dn}(u, \kappa).
\label{J7}
\end{equation}
Following the convention implemented in Maple, we have denoted the second argument by $\kappa$
rather than $\kappa^2$. 

\section{Translationally invariant approach}
For the sake of comparison with the crystal, we also need results for the translational invariant 
HF solution \cite{18}. Let us start with the  vacuum and recall the basic equations \cite{5}. 
The vacuum energy density reads
\begin{eqnarray}
E_0 & = &  - 2 \int_0^{\Lambda/2} \frac{{\rm d}p}{2\pi} \sqrt{p^2+m^2} + \frac{(m-m_0)^2}{2Ng^2}
\nonumber \\
& = & - \frac{\Lambda^2}{8\pi} - \frac{m^2}{4\pi} + \frac{m^2}{2\pi} \ln \frac{m}{\Lambda} + \frac{(m-m_0)^2}{2Ng^2}.
\label{I1}
\end{eqnarray}
Varying with respect to the Fermion mass $m$ yields the 
gap equation 
\begin{equation}
\frac{\pi}{Ng^2} = \gamma +  \ln \frac{\Lambda}{m}
\label{I2}
\end{equation}
with 
\begin{equation}
\gamma = \frac{m_0}{m}\ln \frac{\Lambda}{m}.
\label{I3}
\end{equation}
From now on, we set $m=1$ as in the main text. For the vacuum energy density at the minimum, we get
\begin{equation}
E_0 = - \frac{\Lambda^2}{8\pi} - \frac{1}{4\pi} - \frac{\gamma}{2\pi}.
\label{I4}
\end{equation}
The energy density of the massive Fermi gas differs from Eq.~(\ref{I1}) only in the integration limits,
\begin{eqnarray}
E & = & - 2 \int_{p_f}^{\Lambda/2} \frac{{\rm d}p}{2\pi} \sqrt{p^2+M^2} + \frac{(M-m_0)^2}{2Ng^2}
\nonumber \\
& = & - \frac{\Lambda^2}{8\pi} - \frac{M^2}{4\pi} + \frac{M^2}{2\pi} \ln \frac{M}{\Lambda} + \frac{(M-m_0)^2}{2Ng^2}
\nonumber \\
& &  + \frac{p_f  \epsilon_f}{2\pi} + \frac{M^2}{2\pi} \ln \left( \frac{p_f+\epsilon_f }{M}\right)
\label{I5}
\end{eqnarray}
with
\begin{equation}
\epsilon_f = \sqrt{p_f^2+M^2}
\label{I6}
\end{equation}
Eliminating the bare coupling constant with the help of the vacuum gap equation, the logarithmic divergence
disappears and we are left with
\begin{equation}
E = - \frac{\Lambda^2}{8\pi} - \frac{M^2}{4\pi} + \left(\frac{M^2}{2\pi}-\frac{M}{\pi}\right)\gamma + \frac{p_f \epsilon_f}{2\pi}
+ \frac{M^2}{2\pi} \ln \left(p_f  + \epsilon_f\right),
\label{I7}
\end{equation}
$M$ can be obtained by minimizing $E$ with respect to $M$, 
\begin{equation}
(M-1)\gamma  +  M \ln \left(p_f+\epsilon_f\right)  = 0.
\label{I8}
\end{equation}
The difference between energy density and vacuum energy density is finite,
\begin{equation}
E-E_0 = - \frac{(M^2- 1)}{4\pi}- \frac{(M-1)}{2\pi}\gamma + \frac{p_f \epsilon_f}{2\pi}
\label{I9}
\end{equation}
This last equation is only valid at the minimum, since we have made use of Eq.~(\ref{I8}) to simplify the
expression.

In general, Eq.~(\ref{I8}) can only be solved numerically. In some limiting cases, one can solve it by a series expansion for $M$ 
and compute the corresponding energy. Consider the following limits:
\begin{itemize}
\item $\gamma \to \infty$ at fixed $p_f$ (heavy quark limit), setting $\epsilon_f=\sqrt{1+p_f^2}$:
\begin{eqnarray}
M & \approx &  1 - \frac{1}{\gamma} \ln (p_f+\epsilon_f)
\nonumber \\
& & + \frac{1}{\gamma^2} \left[ \ln^2(p_f+\epsilon_f)+ \frac{\ln(p_f+ \epsilon_f)}
{\epsilon_f(p_f+\epsilon_f)}\right].
\label{I10}
\end{eqnarray}
Energy density (without vacuum subtraction),
\begin{eqnarray}
E & \approx & - \frac{\gamma}{2\pi} - \frac{1}{4\pi} + \frac{p_f \epsilon_f}{2\pi} + \frac{1}{2\pi} \ln(p_f+\epsilon_f)
\nonumber \\ 
& & - \frac{1}{2 \pi \gamma}  \ln^2 (p_f+ \epsilon_f).
\label{I11}
\end{eqnarray}
\item $p_f \to \infty$ at fixed $\gamma$ (high density limit):
\begin{equation}
M \approx \frac{\gamma}{\gamma + \ln 2p_f}- \frac{1}{4p_f^2} \frac{\gamma^3}{(\gamma  + \ln 2p_f)^4},
\label{I12}
\end{equation}
\begin{equation}
E \approx \frac{p_f^2}{2\pi} - \frac{1}{2\pi} \frac{\gamma^2}{\gamma + \ln 2p_f}
+ \frac{1}{16 \pi p_f^2} \frac{\gamma^4}{(\gamma + \ln 2 p_f)^4}.
\label{I13}
\end{equation}
\end{itemize}
As described by Barducci et al.  \cite{18}, the translationally invariant solution undergoes a first order
phase transition at a certain critical chemical potential or density. In the HF solution described above,
this manifests itself through the fact that the energy-versus-$p_f$ curve starts out as concave at $p_f=0$ and
becomes convex only above a certain critical Fermi momentum $p_f^c$. Before drawing conclusions
about the stability of the crystal, we have to take this fact into account.

Just like at $\gamma=0$, we make a mixed phase variational ansatz \cite{9}: A fraction $\lambda$ of space
contains all the extra fermions (``droplets"), which therefore have an increased Fermi momentum 
\begin{equation}
p_f' = \frac{p_f}{\lambda}.
\label{I14}
\end{equation}
The energy density, subtracting the vacuum contribution, is 
\begin{eqnarray}
\Delta  E &=& \lambda \left\{ - \frac{M^2}{4\pi}+ \left( \frac{M^2}{2\pi}-\frac{M}{\pi}\right)\gamma \right.
\label{I15} \\
& & \left. + \frac{p_f'\epsilon_f'}{2\pi} 
+ \frac{M^2}{2\pi} \ln \left(p_f'+\epsilon_f'\right)+ \frac{1}{4\pi}+ \frac{\gamma}{2\pi}\right\}
\nonumber
\end{eqnarray}
with
\begin{equation}
\epsilon_f' =\sqrt{M^2+ (p_f')^2}
\label{I16}
\end{equation}
Vary with respect to $\lambda$,
\begin{equation}
\frac{\partial \Delta  E}{\partial \lambda}=0 \ \ \leftrightarrow \ \ \Delta  E- p_f' \frac{\partial \Delta  E}{\partial p_f'}=0 .
\label{I17}
\end{equation}
The right hand side can be interpreted as construction of the convex hull of the curve energy density versus
fermion density (remember the geometrical meaning of the Legendre transform). The solution $p_f'$ of this equation
 at $\lambda=1$ is  
the critical Fermi momentum. More explicitly, Eq.~(\ref{I17}) reads
\begin{equation}
(1+M+2\gamma-2\gamma M)(1-M)-2 p_f'\epsilon_f' + 2 M^2 \ln \left(p_f'+\epsilon_f'\right)=0.
\label{I18}
\end{equation}
Vary with respect to $M$ [cf.  Eq.~(\ref{I8})],
\begin{equation}
(M-1)\gamma + M \ln \left(p_f'+\epsilon_f'\right)=0.
\label{I19}
\end{equation}
Eqs.~(\ref{I18}) and (\ref{I19}) determine the mixed phase. To solve them, proceed as follows: Eliminate
the ln-term in Eq.~(\ref{I18}) with the help of Eq.~(\ref{I19}) and solve the resulting equation for $p_f'$,
\begin{equation}
(p_f')^2 = \frac{-M^2+ \sqrt{M^4+(1-M)^2(1+M+2\gamma)^2}}{2}.
\label{I20}
\end{equation}
Insert this result into Eq.~(\ref{I19}) and arrive at an equation relating $M$ and $\gamma$. This equation
always has a non-trivial solution, which has to be determined numerically. Analytical results can be
obtained in limiting cases of  interest,
\begin{eqnarray}
M & \approx  & \frac{2}{\ln 2}\gamma - \frac{8}{(\ln 2)^2}\gamma^2
+ \frac{4[10 \ln 2 + (\ln 2)^2 + 2]}{(\ln 2)^4}\gamma^3 
\nonumber \\
& &   {\rm for}\ \gamma \to 0,
\nonumber \\
M & \approx & 1- \frac{3}{2 \gamma^2}+ \frac{3}{\gamma^3}- \frac{63}{40 \gamma^4} \quad {\rm for}\ \gamma \to \infty.
\label{I21}
\end{eqnarray}
By inserting these expressions into Eqs.~(\ref{I20}), (\ref{I15}), we can determine the behavior of the
energy in the mixed phase, as well as the critical point. Find for small $\gamma$
\begin{eqnarray}
\Delta  E& \approx & \frac{p_f}{\pi} \left\{ \frac{1}{\sqrt{2}} + \frac{1}{\sqrt{2}}\gamma - \frac{4+\ln 2}{2\sqrt{2}\ln 2}
 \gamma^2\right\},
\nonumber \\
p_f^c & \approx & \frac{1}{\sqrt{2}} + \frac{1}{\sqrt{2}}\gamma - \frac{8+4 \ln 2 + (\ln 2)^2}{2 \sqrt{2}(\ln 2)^2}\gamma^2,	
\label{I22}
\end{eqnarray}
and for large $\gamma$
\begin{eqnarray}
\Delta  E & \approx & \frac{p_f}{\pi} \left\{1- \frac{3}{8 \gamma^2} + \frac{3}{4\gamma^3}\right\},
\nonumber \\
p_f^c & \approx & \frac{3}{2\gamma} - \frac{3}{2 \gamma^2} - \frac{69}{80 \gamma^3}.
\label{I23}
\end{eqnarray}
The expressions in curly brackets in Eqs.~(\ref{I22}) and (\ref{I23}) are the critical chemical potentials (or,
equivalently, the baryon masses in a kind of ``bag model" \cite{9}).
Compare this with the true baryon masses in the two limits [cf.  Eqs.~(\ref{E8},\ref{E9})] ,
\begin{eqnarray}
M_B & \approx & \frac{2}{\pi} - \frac{\gamma}{\pi}\left(1+ \ln \frac{\gamma}{4} \right) \qquad{\rm for}\ \gamma \to 0,
\nonumber \\
M_B & \approx & 1- \frac{\pi^2}{24 \gamma^2} + \frac{\pi^2}{12 \gamma^3} \qquad {\rm for}\  \gamma \to \infty.
\label{I24}
\end{eqnarray}
In these limits, the baryon mass is below the bag model mass. One can easily check numerically 
that this is in fact true for arbitrary values of $\gamma$. This proves that the translationally invariant
solution is always unstable against formation of a kink-antikink crystal.


\begin{thebibliography}{99}

\bibitem{1}
D. J. Gross and A. Neveu, Phys. Rev. D {\bf 10}, 3235 (1974).
\bibitem{2}
M. Thies and K. Urlichs, Phys. Rev. D {\bf 67}, 125015 (2003).
\bibitem{3}
M. Thies, Phys. Rev. D {\bf 69}, 067703 (2004). 
\bibitem{4}
O. Schnetz, M. Thies and K. Urlichs, Ann. Phys. {\bf 314}, 425 (2004).
\bibitem{5}
M. Thies and K. Urlichs, Phys. Rev. D {\bf 71}, 105008 (2005).
\bibitem{6}
M. J. Rice, S. R. Phillipot, A. R. Bishop, and D. K. Campbell, Phys. Rev. B {\bf 34}, 4139 (1986).
\bibitem{7}
A. Saxena and A. R. Bishop, Phys. Rev. A {\bf 44}, R2251 (1991).
\bibitem{8}
A. Chodos and H. Minakata, Phys. Lett. A {\bf 191}, 39 (1994).
\bibitem{9}
V. Sch\"on and M. Thies, in: {\em At the frontier of particle physics:
Handbook of QCD}, Boris Ioffe Festschrift, ed. by
M. Shifman, Vol. 3, ch. 33, p. 1945, World Scientific, Singapore (2001).
\bibitem{9a}
A. J. Heeger, S. Kivelson, J. R. Schrieffer, and W.-P. Su, Rev. Mod. Phys. {\bf 60}, 781 (1988).
\bibitem{9b}
A. J. Niemi and G. W. Semenoff, Phys. Rep. {\bf 135}, 99 (1986).
\bibitem{10}
R. F. Dashen, B. Hasslacher and A. Neveu, Phys. Rev. D {\bf 12}, 2443 (1975).
\bibitem{11}
S. A. Brazovskii and N. N. Kirova, JETP Lett. {\bf 33}, 4 (1981).
\bibitem{12}
YU Lu, ed., {\em Solitons \& polarons in conducting polymers}, World Scientific, Singapore (1988).
\bibitem{13}
S. A. Brazovskii, N. N. Kirova and S. I. Matveenko, Sov. Phys. JETP {\bf 59}, 434 (1984). 
\bibitem{14}
A. Saxena and J. D. Gunton, Phys. Rev. B {\bf 35}, 3914 (1987).
\bibitem{15}
A. Saxena and J. D. Gunton, Phys. Rev. B {\bf 38}, 8459 (1988).
\bibitem{16}
A. Saxena and W. Cao, Phys. Rev. B {\bf 38}, 7664 (1988).
\bibitem{17}
P. S. Davids, A. Saxena, and D. L. Smith, Phys. Rev. B {\bf 53}, 4823 (1996).
\bibitem{18}
A. Barducci, R. Casalbuoni, M. Modugno, G. Pettini, and R. Gatto, Phys. Rev. D {\bf 51},
3042 (1995).
\bibitem{18a}
R. E. Peierls, {\em Quantum theory of solids}, Clarendon, Oxford (1955), p. 108.
\bibitem{18b}
E. T. Whittaker and G. N. Watson, {\em A course of modern analysis}, Cambridge U. Press (1980).
\bibitem{18c}
H. Li, D. Kusnezov and F. Iachello, J. Phys. A: Math. Gen. {\bf 33}, 6413 (2000).
\bibitem{19}
M. Abramowitz and I. A. Stegun, eds., {\em Handbook of mathematical functions},
Dover Publications, New York (1972), p. 599, Eq.~(17.7.8).
\bibitem{20}
J. Feinberg and A. Zee, Phys. Lett. B {\bf 411}, 134 (1997).
\bibitem{20a}
G. Baym, {\em Lectures on quantum mechanics}, Benjamin/Cummings, Reading,
Massachusetts (1972), p. 237.

\end{thebibliography}
\end{document}